\documentstyle[11pt] {article}

\title{Newton's Third Law in the Framework of Special Relativity for Charged Bodies}

\author{Shailendra Rajput$^a$ \& Asher Yahalom$^{a,b}$ \\
$^a$ Ariel University, Kiryat Hamada POB 3, Ariel 40700, Israel\\
$^b$ Princeton University, Princeton, New Jersey 08543, USA\\
e-mails: shailendrara@ariel.ac.il, asya@ariel.ac.il}

\begin{document}
\maketitle

\newcommand{\beq} {\begin{equation}}
\newcommand{\enq} {\end{equation}}
\newcommand{\ber} {\begin {eqnarray}}
\newcommand{\enr} {\end {eqnarray}}
\newcommand{\eq} {equation}
\newcommand{\eqs} {equations }
\newcommand{\mn}  {{\mu \nu}}
\newcommand{\sn}  {{\sigma \nu}}
\newcommand{\rhm}  {{\rho \mu}}
\newcommand{\sr}  {{\sigma \rho}}
\newcommand{\bh}  {{\bar h}}
\newcommand {\er}[1] {equation (\ref{#1}) }
\newcommand {\ern}[1] {equation (\ref{#1})}
\newcommand {\Ern}[1] {Equation (\ref{#1})}
\newcommand{\curl}[1]{\vec{\nabla} \times #1} % for curl

\begin {abstract}
Newton's third law states that any action is countered by a reaction of equal magnitude but opposite direction. The total force in a system not affected by external forces is thus zero. However, according to the principles of relativity, a signal cannot propagate at speeds exceeding the speed of light. Hence the action and reaction cannot be generated at the same time due to the relativity of simultaneity. Thus, the total force cannot be null at a given time.
In a previous paper \cite{MTAY1}, we have shown that Newt\-on'n third law cannot strictly
hold in a distributed system, where the different parts are at a finite distance from each other. This is due to the finite speed of signal propagation, which cannot exceed the speed of light in the vacuum. A specific example of two current loops with time dependent currents demonstrated that the summing of the total force in the system does not add up to zero. This analysis led to the suggestion of a relativistic engine \cite{MTAY3,AY1}. As the system is affected by a total force for a finite period, the system acquires mechanical momentum and energy. Now the question then arises how can we accommodate the law of momentum and energy conservation. The subject of momentum conversation was discussed in \cite{MTAY4}, while preliminary results regarding energy conservation were discussed in \cite{AY2,RY,RY2}. Previous analysis relied on the fact that the bodies were macroscopically natural, which means that the number of electrons and ions is equal in every volume element. Here we relax this assumption and study charged bodies, thus analyzing the consequences on a possible electric relativistic engine.

\vspace*{0.5 cm}

\vspace*{0.5 cm}
\end {abstract}

\section{Introduction}

Special relativity is a theory of the structure of space-time. It was introduced in Einstein's famous 1905 paper: "On the Electrodynamics of Moving Bodies" \cite{Einstein}. This theory was a consequence of empiric observations and the laws of electromagnetism, which were formulated in the middle of the nineteenth century by Maxwell in his famous four partial differential equations \cite{Maxwell,Jackson,Feynman} which owe their current form to Oliver Heaviside \cite{Heaviside}.
One of the consequences of these equations is that an electromagnetic signal travels at the speed of light $c$, which led people to believe that light is an electromagnetic wave. This was later used by Albert Einstein \cite{Einstein,Jackson,Feynman} to formulate his special theory of relativity, which postulates that the speed of light in vacuum $c$ is the maximal allowed velocity in nature. According to the theory of relativity, any object, message, signal (even if not electromagnetic), or field can not travel faster than the speed of light in vacuum. Hence retardation, if someone at a distance $R$ from me changes something, I may not know about it for at least a retardation time of $\frac{R}{c}$. This means that action and its reaction cannot be generated simultaneously because of the signal finite propagation speed.

Newton's laws of motion are three physical laws that, together, laid the foundation for classical mechanics. These laws describe the relationship between the body, acting forces, and motion of body in response to those forces. The three laws of motion were first compiled by Isaac Newton in his Philosophiae Naturalis Principia Mathematica (Mathematical Principles of Natural Philosophy), first published in 1687 \cite{Newton,Goldstein}. In this paper, we are interested only in the third law, which states: When one body exerts a force on a second body, the second body \textbf{simultaneously} exerts a force equal in magnitude and opposite in direction on the first body.

According to the third law, the total force in a system not affected by external forces is thus zero. This law has numerous experimental verifications and seems to be one of the corner stones of physics. However, in light of the previous discussion it is obvious that action and its reaction cannot be generated at the same time because of the finite speed of signal propagation. Hence the third law is false in an exact sense, although it can be true for most practical applications due to the high speed of signal propagation. Thus the total force cannot be null at a given time.

The locomotive systems of today are based on two material parts; each obtains momentum that is equal and opposite to the momentum gained by the second part. A typical example of this type of system is a rocket that sheds exhaust gas to propel itself. However, the above relativistic considerations suggest a new type of motor in which the system is not composed of two material bodies but of a material body and field. Ignoring the field, a naive observer will see the material body gaining momentum created out of nothing. However, a knowledgeable observer will understand that the opposite amount of momentum is obtained by the field \cite{MTAY4}. Indeed, Noether's theorem dictates that any system that possesses translational symmetry will conserve momentum. The total physical system containing matter and field is indeed symmetrical under translations, while every sub-system (either matter or field) is not.  This was already noticed by Feynman \cite{Feynman}. Feynman describes two orthogonally moving charges, apparently violating Newton's third law as the forces that the charges induce on each other do not cancel (last part of 26-2), this paradox is resolved in (27-6) in which it is shown that the momentum gained by the two charge system is balanced by the field momentum.

 In what follows, we will assume that the magnetization and polarization of the medium are small and therefore we neglect corrections to the Lorentz force suggested in \cite{Mansuripur}. In a paper by Griffiths \&  Heald \cite{Griffiths} it was pointed out that strictly Coulomb�s law and the Biot-Savart law determine the electric and magnetic fields for static sources only. Time-dependent generalizations of these two laws introduced by Jefimenko \cite{Jefimenko} were used to explore the applicability of Coulomb and Biot-Savart outside the static domain.

In a previous paper, we used Jefimenko's \cite{Jefimenko,Jackson} equation to discuss the force between two current carrying coils \cite{MTAY1}. This was later expanded to include the interaction between a current carrying loop and a permanent magnet  \cite{MTAY3,AY1}. Since the system is affected by a total force for a finite period, hence the system acquires mechanical momentum and energy. The question then arises if we need to abandon the law of momentum and energy conservation. The subject of momentum conversation was discussed in \cite{MTAY4}. In \cite{AY2,RY,RY2} some preliminary aspects of the exchange of energy between the mechanical part of the relativistic engine and the electromagnetic field were discussed. In particular, it was shown that the electric energy expenditure is twice the kinetic energy gained by the relativistic motor. It was also shown how some energy might be radiated from the relativistic engine device if the coils are not configured properly.

Previous analysis relied on the fact that the bodies were macroscopically natural, which means that the number of electrons and ions is equal in every volume element. Here we relax this assumption and study charged bodies, thus analyse the consequences on a possible electric relativistic engine.

This paper will use Jefimenko�s equation to discuss the force between two charge carrying bodies. Two mathematical treatments will be given; in one we consider an instantaneous action at a distance, here Newton's third law hold, and the total electromagnetic force will be shown to be equal to zero. Then we consider the dynamic electromagnetic condition where the reaction to an action cannot occur before having the action-generated information reach the affected object, thus bringing about a non-zero resultant. We stress that the bodies themselves are taken to be stationary, it's only the charges in them that are changing with time.

\section{The Electro - Static Condition}
\label{Elst}

\begin{figure}
\vspace{3cm} \includegraphics{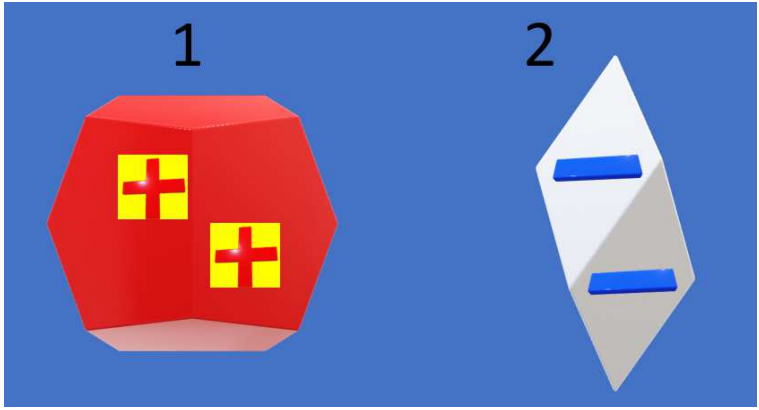}
\caption {Two charged bodies.}
 \label{twoch}
\end{figure}

Consider two bodies having volume elements $d^3 x_1,d^3 x_2$ located at $\vec x_1,\vec x_2$  respectively and
having static charge densities $\rho_1, \rho_2$ (see figure \ref{twoch}).
The electro-static force applied by charged body $2$ on charged body $1$ is given by Jackson \cite{Jackson}:
\beq
\vec F_{12} = \frac{1}{4 \pi \epsilon_0} \int \int  \frac{ d^3 x_1 d^3 x_2 \ \rho_1 (\vec x_1) \rho_2 (\vec x_2) \ \vec x_{12}}{|\vec x_{12}|^3}.
\label{Fd}
\enq
where, $\vec x_{12} =\vec x_1 - \vec x_2$ and $\epsilon_0 =8.85~10^{-12}$ $\rm F m^{-1}$ is the vacuum permittivity.
From the above, it is easy to calculate the electro-static force applied by current loop $1$ on charged body $2$
 by changing indices $1\leftrightarrow2$:
\beq
\vec F_{21} =  \frac{1}{4 \pi \epsilon_0} \int \int  \frac{ d^3 x_1 d^3 x_2 \ \rho_1 (\vec x_1) \rho_2 (\vec x_2) \ \vec x_{21}}{|\vec x_{12}|^3}.
\label{Fe}
\enq
However,
\beq
\vec x_{21} =\vec x_2 - \vec x_1 = -(\vec x_1 - \vec x_2)=-\vec x_{12}, \qquad |\vec x_{21}| = |\vec x_{12}|
\label{x21}
\enq
We obtain:
\beq
\vec F_{21} = -\vec F_{12}.
\label{Ff}
\enq
And the total force on two loop system is null:
\beq
\vec F_T = \vec F_{12} + \vec F_{21} = 0.
\label{Ftotal}
\enq
The above result is independent of the geometry of the charged bodies and is in complete agreement with Newton's third law.

\section{The Dynamic Electromagnetic Condition}

Let us now consider the more general time dependent case. Maxwell's equations dictate that in this case one can not have
a magnetic field without an electric field and vice versa. Therefore we will consider both the electric
and magnetic parts of the Lorenz force $\vec F_{21}$. Let us suppose that the electric field $\vec E$ and magnetic field $\vec B$
 are created by charged body $1$ and acts upon charged body $2$.
Since a charged body may contain both ions and free electrons we will have the Lorentz force:
\beq
\vec F_{21} = \int d^3 x_2 \rho_{i2} (\vec E + \vec v_{i2} \times \vec B) + \int d^3 x_2 \rho_{e2} (\vec E + \vec v_{e2} \times \vec B).
\label{F21ie}
\enq
In the above, we integrate over the entire volume of charged body $2$. $\rho_{i2}$ and $\rho_{e2}$ are the ion charge density and
electron charge density respectively, $\vec v_{i2}$ and $\vec v_{e2}$ are the ion velocity field and
electron velocity field respectively. The total charge density amounts the sum of the ions charge density  and free electrons charge density, hence:
\beq
\rho_{2} = \rho_{i2} + \rho_{e2}.
\label{denie}
\enq
Thus the electric terms in the above force equation cancel and we are left with:
\beq
\vec F_{21} = \int d^3 x_2 \rho_{2} \vec E  + \int d^3 x_2 \rho_{i2} \vec v_{i2} \times \vec B + \int d^3 x_2 \rho_{e2}  \vec v_{e2} \times \vec B.
\label{F21ie2}
\enq
In the laboratory frame the ions being at rest we have: $\vec v_{i2}=0$. Thus, we arrive at:
\beq
\vec F_{21} = \int d^3 x_2 \rho_{2} \vec E  + \int d^3 x_2 \rho_{e2}  \vec v_{e2} \times \vec B.
\label{F21ie3}
\enq
Introducing the current density: $\vec J_2 = \rho_{e2} \vec v_{e2}$, we obtain:
\beq
\vec F_{21} = \int d^3 x_2 \left( \rho_{2} \vec E  + \vec J_2 \times \vec B \right).
\label{F21ie4}
\enq
Now, let us consider the coil that generates the magnetic field. The electric and magnetic fields can be written as follows in terms of the vector and scalar potentials \cite{Jackson}:
\beq
\vec E = - \partial_t \vec A - \vec \nabla \Phi.
\label{E1}
\enq
\beq
\vec B = \vec \nabla \times \vec A.
\label{BA}
\enq
Here, $\vec \nabla$ has the standard definition in vector analysis, $t$ is time and $\partial_t$ is a partial derivative with respect to time. If the field is generated by a charge density $\rho_1$ and current density $\vec J_1$ in charged body $1$, we can solve for the scalar and vector potentials and obtain the result \cite{Jackson}:
\beq
\Phi (\vec x_{2}) =  \frac{1}{4 \pi \epsilon_0} \int d^3 x_1 \frac{\rho_1 (\vec x_1,t_{ret})}{R}, \qquad \vec R \equiv \vec x_{12},
\qquad t_{ret} \equiv t-\frac{R}{c}.
\label{Phi}
\enq
\beq
\vec A (\vec x_{2}) = \frac{\mu_0}{4 \pi} \int d^3 x_1 \frac{\vec J_1 (\vec x_1,t_{ret})}{R}.
\label{A}
\enq
Here, $c = \frac{1}{\sqrt{\epsilon_0 \mu_0}}$ is the speed of light in vacuum. The above solutions satisfy the Lorentz gauge conditions:
\beq
\vec \nabla \cdot \vec A + \frac{1}{c^2} ~\partial_t \Phi = 0
\label{LG}
\enq
Due to the conservation of charge:
\beq
\vec \nabla \cdot \vec J + \partial_t \rho = 0
\label{Chacon}
\enq
(See appendix A). Combining \ern{A} with \ern{BA}, we arrive at the result:
\beq
\vec B (\vec x_{2}) = \vec \nabla_{\vec x_2} \times \vec A(\vec x_{2}) = \frac{\mu_0}{4 \pi} \int d^3 x_1
\vec \nabla_{\vec x_2} \times \left(\frac{\vec J_1 (\vec x_1,t_{ret})}{R}\right).
\label{BA2}
\enq
However, notice that\footnote{We use the notation $\partial_y \equiv \frac{\partial}{\partial y}$.}:
\beq
\vec \nabla_{\vec x_2} \times \left(\frac{\vec J_1 (\vec x_1,t_{ret})}{R}\right)=
\vec \nabla_{\vec x_2} R \times \partial_R \left(\frac{\vec J_1 (\vec x_1,t_{ret})}{R}\right).
\label{BA2a}
\enq
Since:
\beq
\vec \nabla_{\vec x_2} R =-\frac{\vec R}{R}
\label{BA2b}
\enq
And:
\beq
\partial_R \left(\frac{\vec J_1 (\vec x_1,t_{ret})}{R}\right) =
-\frac{\vec J_1 (\vec x_1,t_{ret})}{R^2}-\frac{\partial_t\vec J_1 (\vec x_1,t_{ret})}{R c}.
\label{BA2c}
\enq
Hence:
\beq
\vec \nabla_{\vec x_2} \times \left(\frac{\vec J_1 (\vec x_1,t_{ret})}{R}\right)=
\frac{\vec R}{R^3} \times \left(\vec J_1 (\vec x_1,t_{ret}) + \left(\frac{R}{c}\right) \partial_t\vec J_1 (\vec x_1,t_{ret}) \right).
\label{BA2d}
\enq
Inserting \ern{BA2d} into \ern{BA2}, we arrive at Jefimenko's equations \cite{Jackson,Jefimenko} for the magnetic field:
\beq
\vec B (\vec x_{2}) =  \frac{\mu_0}{4 \pi} \int d^3 x_1
\frac{\vec R}{R^3} \times \left(\vec J_1 (\vec x_1,t_{ret}) + \left(\frac{R}{c}\right) \partial_t\vec J_1 (\vec x_1,t_{ret}) \right).
\label{BA3}
\enq
For the electric field, we have two contributions according to \ern{E1}, one from the scalar potential and a second one from the vector potential:
\beq
\vec E  =  \vec E_a  + \vec E_b, \qquad  \vec E_a \equiv- \partial_t \vec A, \qquad  \vec E_b \equiv- \vec \nabla \Phi.
\label{E2}
\enq
Hence according to \ern{A}:
\beq
 \vec E_a (\vec x_{2}) = -\frac{\mu_0}{4 \pi} \int d^3 x_1 \frac{\partial_t \vec J_1 (\vec x_1,t_{ret})}{R}.
\label{E3}
\enq
And according to \ern{Phi}:
\beq
\vec E_b (\vec x_{2}) =  - k \int d^3 x_1 \vec \nabla_{\vec x_2} \left(\frac{\rho_1 (\vec x_1,t_{ret})}{R}\right),
\qquad k \equiv \frac{1}{4 \pi \epsilon_0} \simeq 9 \cdot 10^9.
\label{E4}
\enq
Above equation can also be written as:
\beq
\vec E_b (\vec x_{2}) =  - k \int d^3 x_1  \left[\frac{1}{R}\vec \nabla_{\vec x_2}  \rho_1 (\vec x_1,t_{ret})
+ \rho_1 (\vec x_1,t_{ret})  \vec \nabla_{\vec x_2}\frac{1}{R} \right],
\label{E5}
\enq
however:
\beq
\vec \nabla_{\vec x_2}\frac{1}{R} = \frac{\hat R}{R^2}, \qquad \hat R \equiv \frac{\vec R}{R}
\label{E6}
\enq
and also:
\beq
\vec \nabla_{\vec x_2}  \rho_1 (\vec x_1,t_{ret}) = \vec \nabla_{\vec x_2} R~\left. \partial_R \rho_1 (\vec x_1,t_{ret})\right|_{\vec x_1}
= \frac{\hat R}{c} \partial_t \rho_1 (\vec x_1,t_{ret}).
\label{E7}
\enq
It thus follows that:
\beq
\vec E_b (\vec x_{2}) =  - k \int d^3 x_1  \frac{\hat R}{R^2} \left[ \rho_1 (\vec x_1,t_{ret})+ \left(\frac{ R}{c}\right) \partial_t \rho_1 (\vec x_1,t_{ret})\right].
\label{E8}
\enq
Adding $\vec E_b$ and $\vec E_a$ and taking into account that:
\beq
\frac{\mu_0}{4 \pi k} = \mu_0 \epsilon_0 = \frac{1}{c^2}
\label{E9}
\enq
We arrive at the Jefimenko's expression \cite{Jackson,Jefimenko} for the electric field:
\ber
\vec E (\vec x_{2}) &=&  - k \int d^3 x_1  \frac{1}{R^2} \left[
\left(\rho_1 (\vec x_1,t_{ret})+ \left(\frac{ R}{c}\right) \partial_t \rho_1 (\vec x_1,t_{ret})\right) \hat R  \right.
\nonumber \\
 &+& \left. \left(\frac{ R}{c}\right)^2 \frac{\partial_t \vec J_1 (\vec x_1,t_{ret})}{R} \right].
\label{E10}
\enr
Inserting  \ern{BA3} and \ern{E10} into \ern{F21ie4},we arrive at a somewhat lengthy but straightforward expression:
\ber
\vec F_{21} &=&  - k \int \int d^3 x_1 d^3 x_2 \frac{1}{R^2}
\nonumber \\
 & & \hspace{-2cm} \left\{ \rho_2 (\vec x_2,t)  \left[ \left(\rho_1 (\vec x_1,t_{ret})+ \left(\frac{ R}{c}\right) \partial_t \rho_1 (\vec x_1,t_{ret})\right) \hat R + \left(\frac{ R}{c}\right)^2 \frac{\partial_t \vec J_1 (\vec x_1,t_{ret})}{R}  \right]  \right.
 \nonumber \\
 & & \hspace{-2cm} \left. + \left(\frac{ R}{c}\right)^2 \left[\frac{\hat R}{R^2} \times \left(\vec J_1 (\vec x_1,t_{ret}) + \left(\frac{R}{c}\right) \partial_t\vec J_1 (\vec x_1,t_{ret})\right) \right]
 \times \vec J_2 (\vec x_2,t) \right\}
\label{F216}
\enr

\subsection{The Quasi-Static Approximation}
\label{QSA}

In the quasi-static approximation, it is assumed that $\tau=\frac{R}{c}$ is small and can be neglected. Under the same
approximation $t_{ret} \simeq t$. Neglecting all terms of order $\tau$, \ern{F216} takes the Coulomb form:
\beq
\vec F_{21} = - k \int \int  d^3 x_1 d^3 x_2 \ \rho_1 (\vec x_1,t) \rho_2 (\vec x_2,t) \frac{\hat R}{R^2} .
\label{F21qs}
\enq
Above equation is similar to the \ern{Fe} with slightly different notation, but now the charge densities are time dependent and therefore not static.
Newton third law follows trivially using the same arguments as in section \ref{Elst}.
We notice that for vanishing charge densities \ern{F216} takes the form:
\ber
\vec F_{21} &=&  - k \int \int d^3 x_1 d^3 x_2 \frac{1}{R^2}
\nonumber \\
 & & \hspace{-2cm} \left\{   \left(\frac{ R}{c}\right)^2 \left[\frac{\hat R}{R^2} \times \left(\vec J_1 (\vec x_1,t_{ret}) + \left(\frac{R}{c}\right) \partial_t\vec J_1 (\vec x_1,t_{ret})\right) \right]
 \times \vec J_2 (\vec x_2,t) \right\}.
\label{F217}
\enr
In this case the leading term is $O(\tau^2)$, neglecting higher order terms in $\tau$ this can be written as:
\beq
\vec F_{21} =   \frac{\mu_0}{4 \pi} \int \int d^3 x_1 d^3 x_2  \left\{\vec J_2 (\vec x_2,t) \times \left[\frac{\hat R}{R^2} \times \vec J_1 (\vec x_1,t)  \right] \right\}.
\label{F218}
\enq
Or also as:
\ber
\vec F_{21} &=&   \frac{\mu_0}{4 \pi} \int \int d^3 x_1 d^3 x_2  \left\{
 \left(\vec J_2 (\vec x_2,t)\cdot \vec J_1 (\vec x_1,t) \right) \frac{\hat R}{R^2} \right.
 \nonumber \\
&-&  \left. \left(\vec J_2 (\vec x_2,t)\cdot \frac{\hat R}{R^2} \right) \vec J_1 (\vec x_1,t) \right\}.
\label{F219}
\enr
Notice that according to \ern{E6}:
\beq
\vec J_2 (\vec x_2,t)\cdot \frac{\hat R}{R^2} = \vec J_2 (\vec x_2,t)\cdot \vec \nabla_{\vec x_2}\frac{1}{R} =
 \vec \nabla_{\vec x_2} \cdot \left(\frac{\vec J_2 (\vec x_2,t)}{R}\right) - \frac{1}{R} \vec \nabla_{\vec x_2} \cdot \vec J_2  (\vec x_2,t).
\label{F219b}
\enq
However, since we assume that the charge density is null it follows from \ern{Chacon} that $\vec \nabla_{\vec x_2} \cdot \vec J_2 = 0$ and:
\beq
\vec J_2 (\vec x_2,t)\cdot \frac{\hat R}{R^2} =  \vec \nabla_{\vec x_2} \cdot \left(\frac{\vec J_2 (\vec x_2,t)}{R}\right).
\label{F219c}
\enq
Takin the volume integral of the left hand side and using Gauss theorem, we obtain:
\beq
 \int d^3 x_2 ~\vec J_2 (\vec x_2,t)\cdot \frac{\hat R}{R^2} =  \oint d \vec S_2 \cdot \left(\frac{\vec J_2 (\vec x_2,t)}{R}\right),
\label{F219d}
\enq
where $d \vec S_2$ and the integral is taken over the entire surface encapsulating the volume. If the surface is take far from the system in which the current density is non-vanishing it follows that the surface integral is null and thus:
\beq
 \int d^3 x_2 ~\vec J_2 (\vec x_2,t)\cdot \frac{\hat R}{R^2} = 0.
\label{F219e}
\enq
Thus \ern{F219} will take the form:
\beq
\vec F_{21} =   \frac{\mu_0}{4 \pi} \int \int d^3 x_1 d^3 x_2
 \left(\vec J_2 (\vec x_2,t)\cdot \vec J_1 (\vec x_1,t) \right) \frac{\hat R}{R^2} .
\label{F2110}
\enq
From which Newton's third law follows trivially. We notice that this form is a generalization of a similar result presented in \cite{MTAY1} for the case of currents that are restricted to flow on thin current loops. In the quasi-static approximation, Newton's third law holds regardless of the geometry of the system involved, and the total force is null.

This should come as no surprise to the reader since, in the quasi-static approximation, we totally neglect the time it takes a signal to propagate from system $1$ to system $2$, assuming that this time is zero. Different results are obtained when $\tau$ is not neglected, as we will see in the next subsection.

\subsection{The Case of a Finite $\tau$}

Consider the charge density $\rho(\vec x', t_{ret}) = \rho(\vec x',t - \frac{R}{c})$, if $\frac{R}{c}$ is small but not zero one can write a Taylor series expansion around $t$ in the form:
\beq
\rho(\vec x', t_{ret}) = \rho(\vec x',t - \frac{R}{c})=\sum_{n=0}^{\infty} \frac{\partial^n_t \rho (\vec x',t)}{n!}(- \frac{R}{c})^n.
\label{TayI}
\enq
In the above $\partial^n_t \rho (\vec x',t)$ is the partial temporal derivative of order $n$ of $\rho(\vec x', t)$. The same expansion
for the current density leads to the expression:
\beq
\vec J (\vec x', t_{ret}) = \vec J(\vec x',t - \frac{R}{c})=\sum_{n=0}^{\infty} \frac{\partial^n_t \vec J(\vec x',t)}{n!}(- \frac{R}{c})^n.
\label{TayIb}
\enq
In the above $\partial^n_t \vec J(\vec x',t)$ is the partial temporal derivative of order $n$ of $\vec J(\vec x', t)$.
The above  expansions are valid only for a certain environment of $t$ on the time axis which depends on the functions involved, this environment shall be defined using the convergence radius $T_{max}$ which may be different for each function. That is \ern{TayI} and \ern{TayIb} is valid only in the domain $[t-T_{max},t+T_{max}]$. As we expand in the delay time $\frac{R}{c}$ it follows that the expansion is valid only for a limited range:
\beq
R < R_{max} \equiv c ~ T_{max}.
\label{Rmax}
\enq
This means that basically we are dealing with a near field approximation, however, as $c$ is a large number $R_{max}$ would be quite large
for most systems. Now inserting \ern{TayI} into \ern{Phi} we obtain:
\ber
\Phi_1 (\vec x_{2}) &=& k \int d^3 x_1 \frac{\rho_1 (t_{ret})}{R}=
k \sum_{n=0}^{\infty} \frac{1}{n!} \int d^3 x_1 \partial^n_t \rho_1 (\vec x',t) \frac{1}{R} (- \frac{R}{c})^n
\nonumber \\
&=& k \sum_{n=0}^{\infty} \frac{1}{n!} \left(-\frac{1}{c}\right)^n \int d^3 x_1 \partial^n_t \rho_1 (\vec x_1,t) R^{n-1}
\label{phie2}
\enr
Similarly inserting \ern{TayIb} into \ern{A} we obtain:
\ber
\vec A_1 (\vec x_{2}) &=& \frac{\mu_0}{4 \pi} \int d^3 x_1 \frac{\vec J_1 (t_{ret})}{R}=
\frac{\mu_0}{4 \pi}\sum_{n=0}^{\infty} \frac{1}{n!} \int d^3 x_1 \partial^n_t \vec J_1 (\vec x',t) \frac{1}{R} (- \frac{R}{c})^n
\nonumber \\
&=& \frac{\mu_0}{4 \pi} \sum_{n=0}^{\infty} \frac{1}{n!} \left(-\frac{1}{c}\right)^n \int d^3 x_1 \partial^n_t \vec J_1 (\vec x_1,t) R^{n-1}
\label{Ae2}
\enr
As $\frac{\mu_0}{4 \pi} = \frac{\mu_0 \epsilon_0}{4 \pi \epsilon_0} =\frac{k}{c^2}$ it follows that:
\ber
\vec A_1 (\vec x_{2}) &=& k \sum_{n=0}^{\infty} \frac{1}{n!} \left(-\frac{1}{c}\right)^{n+2} \int d^3 x_1 \partial^n_t \vec J_1 (\vec x_1,t) R^{n-1}
\nonumber \\
&=& k \sum_{n=2}^{\infty} \frac{1}{(n-2)!} \left(-\frac{1}{c}\right)^n \int d^3 x_1 \partial^{n-2}_t \vec J_1 (\vec x_1,t) R^{n-3}
\label{Ae3}
\enr
We are now at a position in which we can calculated the electric and magnetic fields from the expansions given in \ern{phie2} and \ern{Ae3}. However, before we proceed we introduce the following notation. Let $G^{[n]}$ be the contribution of order $\frac{1}{c^n}$ to the quantity $G$, thus:
\beq
G = \sum_{n=0}^{\infty} G^{[n]}
\label{Gn}
\enq
Hence:
\beq
\Phi_1^{[n]} (\vec x_{2}) =  \frac{k}{n!} \left(-\frac{1}{c}\right)^n \int d^3 x_1 \partial^n_t \rho_1 (\vec x_1,t) R^{n-1}
\label{phie2n}
\enq
\ber
&& \vec A_1^{[n]} (\vec x_{2}) =  \frac{k}{(n-2)!} \left(-\frac{1}{c}\right)^n \int d^3 x_1 \partial^{n-2}_t \vec J_1 (\vec x_1,t) R^{n-3} \quad {\rm for} ~n>=2
\nonumber \\
&& \vec A_1^{[0]} = \vec A_1^{[1]} = 0.
\label{Ae3n}
\enr
It now follows from \ern{E2} that:
\ber
&& \vec E_{1a}^{[n]} = - \partial_t \vec A_1^{[n]} =
\nonumber \\
&&-  \frac{k}{(n-2)!} \left(-\frac{1}{c}\right)^n \int d^3 x_1 \partial^{n-1}_t \vec J_1 (\vec x_1,t) R^{n-3} \quad {\rm for} ~n>=2
\nonumber \\
&& \vec E_{1a}^{[0]} = \vec E_{1a}^{[1]} = 0.
\label{Ean}
\enr
and
\beq
 \vec E_{1b}^{[n]} = - \vec \nabla_{\vec x_2} \Phi^{[n]} =
-\frac{k}{n!} \left(-\frac{1}{c}\right)^n \int d^3 x_1 \partial^n_t \rho_1 (\vec x_1,t) \vec \nabla_{\vec x_2} R^{n-1} .
\label{Ebn}
\enq
However, since:
\beq
 \vec \nabla_{\vec x_2} R^{n-1} = \vec \nabla_{\vec x_2} R ~\partial_R R^{n-1} = - (n-1) R^{n-2} \hat R.
\label{Rnd}
\enq
it follows that:
\beq
 \vec E_{1b}^{[n]} =
\frac{k(n-1)}{n!} \left(-\frac{1}{c}\right)^n \int d^3 x_1 \partial^n_t \rho_1 (\vec x_1,t) R^{n-2} \hat R .
\label{Ebn2}
\enq
 Zeroth order contribution comes only from the potential part of the electric field and is the Coulomb contribution:
\beq
 \vec E_{1}^{[0]}= \vec E_{1b}^{[0]} = -k \int d^3 x_1 \frac{\rho_1 (\vec x_1,t)} {R^{2} }\hat R
\label{En0}
\enq
We also deduce that $\vec E_{1b}^{[1]} = 0$, hence:
\beq
 \vec E_{1}^{[1]}=\vec E_{1a}^{[1]} + \vec E_{1b}^{[1]} = 0
\label{En1}
\enq
thus there is no first order correction to the electric field in a charged system, first order corrections are also absent in an uncharged system \cite{MTAY1}. The first term containing contributions from both the scalar and vector potentials to the electric field is the second order term:
\beq
 \vec E_{1}^{[2]}=\vec E_{1a}^{[2]} + \vec E_{1b}^{[2]} =  k \left(\frac{1}{c}\right)^2 \int d^3 x_1
  \left[\frac{1}{2}\partial^2_t \rho_1 (\vec x_1,t)\hat R - \partial_t \vec J_1 (\vec x_1,t) R^{-1} \right]
\label{En2}
\enq
As $\frac{1}{c}$ is quite small, it will suffice to consider contributions till the second order. We now calculate the magnetic field
using \ern{BA2} and \ern{Ae3n} and obtain:
\ber
& & \vec B_1^{[n]} (\vec x_{2}) = \vec \nabla_{\vec x_2} \times \vec A_1^{[n]} (\vec x_{2}) =
\nonumber \\
& & \frac{k}{(n-2)!} \left(-\frac{1}{c}\right)^n \int d^3 x_1
\vec \nabla_{\vec x_2} \times \left[\partial^{n-2}_t \vec J_1 (\vec x_1,t) R^{n-3}\right] \quad {\rm for} ~n>=2
\nonumber \\
&& \vec B_1^{[0]} = \vec B_1^{[1]} = 0.
\label{Bn}
\enr
However:
\beq
\vec \nabla_{\vec x_2} \times \left[\partial^{n-2}_t \vec J_1 (\vec x_1,t) R^{n-3}\right] =
\vec \nabla_{\vec x_2} R^{n-3} \times \partial^{n-2}_t \vec J_1 (\vec x_1,t)
\label{Bna}
\enq
and also:
\beq
\vec \nabla_{\vec x_2} R^{n-3} =  \vec \nabla_{\vec x_2} R ~ \partial_R R^{n-3} =
(3-n) ~ R^{n-4} ~\hat R
\label{Bnb}
\enq
Hence we may write:
\ber
& & \hspace{-1cm} \vec B_1^{[n]} (\vec x_{2}) =
\nonumber \\
& & \hspace{-1cm} k \frac{(3-n)}{(n-2)!} \left(-\frac{1}{c}\right)^n \int d^3 x_1
R^{n-4} ~\hat R \times \partial^{n-2}_t \vec J_1 (\vec x_1,t) \quad {\rm for} ~n>=2.
\label{Bn2}
\enr
in particular:
\beq
\vec B_1^{[2]} (\vec x_{2}) =
 k  \left(\frac{1}{c}\right)^2 \int d^3 x_1 R^{-2} ~\hat R \times  \vec J_1 (\vec x_1,t) .
\label{Bn2b}
\enq
Using the expressions of the electric and magnetic fields, we may now calculate the force $n^{th}$ order contribution through \ern{F21ie4}:
\beq
\vec F_{21}^{[n]} = \int d^3 x_2 \left( \rho_{2} (\vec x_{2}) \vec E_1^{[n]} (\vec x_{2})  + \vec J_2 (\vec x_{2}) \times \vec B_1^{[n]} (\vec x_{2})\right).
\label{F21n}
\enq
It follows that for the zeroth order we obtain:
\beq
\vec F_{21}^{[0]} = \int d^3 x_2 \rho_{2} (\vec x_{2}) \vec E_1^{[0]} (\vec x_{2}).
\label{F210}
\enq
Inserting \ern{En0}, we obtain Coulomb's force:
\beq
\vec F_{21}^{[0]} = -k \int  \int d^3 x_1 d^3 x_2 \rho_1 (\vec x_1) \rho_{2} (\vec x_{2}) \frac{1} {R^{2} }\hat R = - \vec F_{12}^{[0]}.
\label{F210b}
\enq
This type of force which is just the quasi static force satisfies Newton's third law (see section \ref{QSA}), hence the total force on the system is null:
\beq
\vec F_T^{[0]} = \vec F_{21}^{[0]} + \vec F_{12}^{[0]} = 0 .
\label{F210c}
\enq
The first order force in $\frac{1}{c}$ is null since the first order electric and magnetic fields are null, thus:
\beq
\vec F_T^{[1]} = \vec F_{21}^{[1]} = \vec F_{12}^{[1]} = 0 .
\label{F211}
\enq
We shall now proceed with the calculation of the second order force term, this will suffice as $\frac{1}{c}$ is a rather small number. To do this we first divide the force given in \ern{F21n} into electric and magnetic terms:
\ber
\vec F_{21}^{[2]} &=& \vec F_{21e}^{[2]} + \vec F_{21m}^{[2]}
\nonumber \\
\vec F_{21e}^{[2]} &\equiv& \int d^3 x_2 ~ \rho_{2} (\vec x_{2}) \vec E_1^{[2]} (\vec x_{2})
\nonumber \\
\vec F_{21m}^{[2]} &\equiv& \int d^3 x_2 ~ \vec J_2 (\vec x_{2}) \times \vec B_1^{[2]} (\vec x_{2})
\label{F212}
\enr
Inserting \ern{En2} into \ern{F212}, we readily obtain the electric force:
\beq
\vec F_{21e}^{[2]} =  \left(\frac{k}{c^2}\right) \int \int d^3 x_1  d^3 x_2 ~
  \left[\frac{1}{2}\rho_2 \partial^2_t \rho_1 \hat R - \rho_2 \partial_t \vec J_1  R^{-1} \right]
\label{F212e}
\enq
The magnetic force can be obtained by inserting \ern{Bn2b} into \ern{F212}:
\ber
\vec F_{21m}^{[2]} &=& \left(\frac{k}{c^2}\right) \int \int d^3 x_1  d^3 x_2 ~ \vec J_2  \times
 ~\left( R^{-2} \hat R \times  \vec J_1  \right)
 \nonumber \\
 &=& \left(\frac{k}{c^2}\right)  \int \int d^3 x_1  d^3 x_2 \left[ \hat R \frac{\vec J_1 \cdot \vec J_2}{R^2}  -
\vec J_1 R^{-2}( \hat R \cdot \vec J_2 )  \right]
\label{F212m}
\enr
However,
\beq
 R^{-2} \hat R = \vec \nabla_{\vec x_2}  R^{-1}
\label{F212mb}
\enq
it follows that:
\beq
\vec F_{21m}^{[2]} =\left(\frac{k}{c^2}\right)  \int \int d^3 x_1  d^3 x_2 \left[ \hat R \frac{\vec J_1 \cdot \vec J_2}{R^2}  -
\vec J_1 \left( (\vec \nabla_{\vec x_2}  R^{-1} ) \cdot \vec J_2 \right)  \right]
\label{F212mc}
\enq
Let us look at the integral:
\beq
\int   d^3 x_2 (\vec \nabla_{\vec x_2}  R^{-1} ) \cdot \vec J_2  =
\int   d^3 x_2 \left[  \vec \nabla_{\vec x_2}  \cdot   \left( \frac{\vec J_2}{R} \right) - \frac{1}{R} \vec \nabla_{\vec x_2}  \cdot \vec J_2 \right]
\label{F212md}
\enq
Using Gauss theorem and the continuity \ern{Chacon}, we arrive at the following expression:
\beq
\int   d^3 x_2 (\vec \nabla_{\vec x_2}  R^{-1} ) \cdot \vec J_2  =
\oint   d \vec S_2  \cdot   \left( \frac{\vec J_2}{R} \right) + \int   d^3 x_2  \frac{1}{R} \partial_t \rho_2.
\label{F212me}
\enq
The surface integral is performed over a surface encapsulating the volume of the volume integral, if the volume integral is performed over all space the surface is at infinity. Provided there are no currents at infinity:
\beq
\int   d^3 x_2 (\vec \nabla_{\vec x_2}  R^{-1} ) \cdot \vec J_2  =
\int   d^3 x_2  \frac{1}{R} \partial_t \rho_2.
\label{F212mf}
\enq
Inserting the result given in \ern{F212mf} into \ern{F212mc}, we arrive at the following expression for the second order magnetic force:
\beq
\vec F_{21m}^{[2]} =\left(\frac{k}{c^2}\right)  \int \int d^3 x_1  d^3 x_2 \left[ \hat R \frac{\vec J_1 \cdot \vec J_2}{R^2}  -
 \frac{\vec J_1}{R} \partial_t \rho_2  \right].
\label{F212mg}
\enq
We remark that for the magnetic field, the second order is the lowest order for the force as the zeroth and first order are null. Moreover, we observe that the force is a sum of two parts, the first one satisfies Newton's third law (see section \ref{QSA}), and the second does not.
We can now calculate the total electromagnetic force by adding \ern{F212e} with \ern{F212mg}:
\ber
\vec F_{21}^{[2]} &=& \vec F_{21e}^{[2]}+ \vec F_{21m}^{[2]}
\nonumber \\
  & & \hspace{-2cm} =\left(\frac{k}{c^2}\right) \int \int d^3 x_1  d^3 x_2 ~
  \left[\frac{1}{2}\rho_2 \partial^2_t \rho_1 \hat x_{12} - \partial_t (\rho_2  \vec J_1)  R^{-1} + \hat x_{12} \frac{\vec J_1 \cdot \vec J_2}{R^2} \right]
\label{F212em1}
\enr
We now use the notation $\hat R = \hat x_{12}$ for clarity. From the above expressions it is easy to calculate $F_{12}^{[2]}$ by exchanging the indices $1$ and $2$:
\ber
\vec F_{12}^{[2]} &=&
\nonumber \\
  & & \hspace{-2cm} \left(\frac{k}{c^2}\right) \int \int d^3 x_1  d^3 x_2 ~
  \left[\frac{1}{2}\rho_1 \partial^2_t \rho_2 \hat x_{21} - \partial_t (\rho_1  \vec J_2)  R^{-1} + \hat x_{21} \frac{\vec J_1 \cdot \vec J_2}{R^2} \right]
\label{F212em2}
\enr
Combining \ern{F212em1} and \ern{F212em2} and taking into account that $\hat x_{12} = - \hat x_{21}$, it follows that:
\ber
\vec F_{T}^{[2]} &=& \vec F_{12}^{[2]} + \vec F_{21}^{[2]} =
\nonumber \\
  & & \hspace{-2cm} \left(\frac{k}{c^2}\right) \int \int d^3 x_1  d^3 x_2 ~
  \left[\frac{1}{2}\left(\rho_2 \partial^2_t \rho_1 - \rho_1 \partial^2_t \rho_2 \right)\hat R - \partial_t (\rho_1  \vec J_2 + \rho_2  \vec J_1)  R^{-1}  \right]
\label{F212t}
\enr
Now as $\frac{k}{c^2} = \frac{\mu_0}{4 \pi}$ and since:
\beq
\rho_2 \partial^2_t \rho_1 - \rho_1 \partial^2_t \rho_2 = \partial_t \left(\rho_2 \partial_t \rho_1 - \rho_1 \partial_t \rho_2 \right)
\label{F212t2}
\enq
It follows that:
\beq
\vec F_{T}^{[2]}= \frac{\mu_0}{4 \pi} \partial_t \int \int d^3 x_1  d^3 x_2 ~
  \left[\frac{1}{2}\left(\rho_2 \partial_t \rho_1 - \rho_1 \partial_t \rho_2 \right)\hat R - (\rho_1  \vec J_2 + \rho_2  \vec J_1)  R^{-1}  \right]
\label{F212t3}
\enq
In the following section, we describe some of the implications of the above formula. We remark that in some fast changing systems the second order correction will not suffice and higher order terms will be needed.

\subsection{Some preliminary observations}

According to Newton's second law, a system with a non zero total force in its center of mass, must have a change in its total linear momentum $\vec P (t)$:
\beq
\vec F_{T}^{[2]} = \frac{d \vec P}{dt}
\label{P}
\enq
Assuming that $\vec P (-\infty) = 0$ and also that there are not current or charge densities at $t=-\infty$, it follows from \ern{F212t3} that:
\beq
\vec P (t)= \frac{\mu_0}{4 \pi}\int \int d^3 x_1  d^3 x_2 ~
  \left[\frac{1}{2}\left(\rho_2 \partial_t \rho_1 - \rho_1 \partial_t \rho_2 \right)\hat R - (\rho_1  \vec J_2 + \rho_2  \vec J_1)  R^{-1}  \right]
\label{P2}
\enq
Comparing \ern{P2} with the momentum gain of a non charged relativistic motor described by equation (64) of \cite{MTAY4}:
\beq
\vec P_{mech} \cong \frac{\mu_0}{8 \pi} \partial_t I_1 (t) I_2 (\frac{h}{c})^2  \vec K_{122}, \qquad  \vec K_{122}=
- \frac{1}{h^2}  \oint \oint \hat R (d\vec l_2 \cdot d\vec l_1)
\label{Pmech1b}
\enq
where $h$ is a typical scale of the system, $I_1 (t)$ and  $I_2$ are the currents flowing through two current loops and
$d\vec l_1, d\vec l_2$ are the current loop line elements. We notice some major differences. First,  we notice a factor of $\left(\frac{h}{c}\right)^2$ in case of an uncharged motor. As, for any practical system the scale $h$ is of the order of one, this means that the charged relativistic motor is stronger than the uncharged motor by a factor of $c^2 \sim 10^{17}$ which is quite a considerable factor. Second, we notice that for the uncharged motor, the current must be continuously increased in order to maintain the momentum in the same direction. Of course, one cannot do this for ever, hence the uncharged motor is a type of a piston engine doing a periodic motion backward and forward and can only produce motion forward by interacting with an external system (the road). This is not the case for the charged relativistic motor. In fact we obtain non vanishing momentum for stationary charge and current densities:
\beq
\vec P (t)= - \frac{\mu_0}{4 \pi} \int \int d^3 x_1  d^3 x_2 ~ (\rho_1  \vec J_2 + \rho_2  \vec J_1)  R^{-1}
\label{P3}
\enq
hence the charged relativistic motor can produce forward momentum without interacting with any external system except the electromagnetic field.
The above expression can be somewhat simplified using the potential given in \ern{Phi} such that:
\beq
\vec P (t)= - \frac{1}{c^2} \left[ \int   d^3 x_2 ~ \Phi_1  \vec J_2 + \int   d^3 x_1 ~ \Phi_2  \vec J_1  \right]
\label{P3b}
\enq
in which we are reminded that $ \frac{\mu_0}{4 \pi k} =  \mu_0 \epsilon_0 = \frac{1}{c^2}$.
Another observation is that in a charged relativistic motor we do not need both subsystems to be charged, in fact if $\rho_2 = 0$:
\beq
\vec P (t)= - \frac{\mu_0}{4 \pi} \int \int d^3 x_1  d^3 x_2 ~ \rho_1  \vec J_2  R^{-1} =
- \frac{1}{c^2} \int   d^3 x_2 ~ \Phi_1  \vec J_2
\label{P4}
\enq
provided that the system has a non vanishing current density $\vec J_2$. We underline that as in \cite{MTAY4} the forward momentum gained
by the mechanical system will be balanced by a backward momentum gained by the electromagnetic system. We shall now discuss the challenges
in developing a charged relativistic motor.

\subsection{Charge density}
\label{chden}
The amount of charge that can be accumulated in a given volume or surface is limited by the phenomena of electrical breakdown in which the surrounding medium is separated into electron and ions and becomes a plasma. As the surrounding becomes conductive a discharge occurs and the charge density is reduced. The dielectric strength $ E_{max}$ of air is 3 MV/m \cite{factbook}, for high vacuum 20-40 MV/m \cite{vacuumds} and for diamond 2000 MV/m \cite{diamondds}. For an infinite surface the surface density $\sigma$ is:
\beq
\sigma=2 \epsilon E <\sigma_{max}= 2 \epsilon E_{max}, \qquad  \epsilon = \epsilon_r \epsilon_0
\label{sigmax}
\enq
in which  $\epsilon_r$ is the relative susceptibility. For air $\sigma_{max} \simeq 53 ~ \mu C /m^2$. To estimate the amount
of charge $Q$ which one can maintain in a given volume we notice that for a spherical symmetric charge ball we obtain at a distance $r$
the radial field $E$:
\beq
E = \frac{k Q}{r^2}
\label{Esph}
\enq
the stronger field is on the ball itself, that is at $r=r_s$ we must have:
\beq
\frac{k Q}{r_s^2} < E_m \Rightarrow Q < Q_m = \frac{1}{k} r_s^2 E_m
\label{Esph2}
\enq
For a typical size of $1$ m, the maximal charge is $3.3 ~ 10^{-4}$ C. Hence regardless if we have surface charge or a volume charge the maximal charge scales as the square of the dimension of the system�, that is as $h^2$. A possible approach to increase the available charge density is to use electret. Fluorinated parylene (Parylene HT, SCS) offers excellent surface charge density of $3.7 ~ {\rm mC/m^2}$ found for a $7.3 ~ {\rm\mu m}$ film by Hsi-wen and Yu-Chong \cite{Hsi}, this material has a dielectric strength of 204.58 MV/m. However, as the thickness of the material grows the charge density is reduced.

\subsection{Current density}
\label{cuden}

The amount of current a device can generate dependent on its voltage and internal resistance. Provided that the external impedance is not significant, the resulting currents are denoted as short currents be the order of a few thousand amperes for a standard domestic electrical installation and as high as hundreds of thousands of amperes in large industrial power systems. If the current is flowing through a metal conductor, then heat is generated due to the finite resistivity of the conductor. Large currents require a thick conductor to avoid excessive heating. This problem can be circumvented using a superconductor, although this will require cooling to extremely low temperatures, making the system quite cumbersome. But, even a superconductor has a critical current density and superconduction properties losses above that. Jung, S. G. et al. \cite{Jung} have reported critical current densities as high as $5~ {\rm kA/cm^2}$. Coil windings will enable to reuse the current, and the number of windings in a given area is also critical to the performance of the system. Proximity of the current to the charge will also affect the amount of generated momenta \er{P4}. However, putting a conductor too close to the charge may result in discharge, hence a balance should be striked.

\subsection{Scalability}

It is clear from \er{P4} that the larger the relativistic engine the more powerful it is. However, since the charge surface density is limited the momenta generate will scale as the second power of the size of the charged subsystem and as the third power of the size of the current carrying system.

\section{An Example}

\begin{figure}
\vspace{7cm} \includegraphics{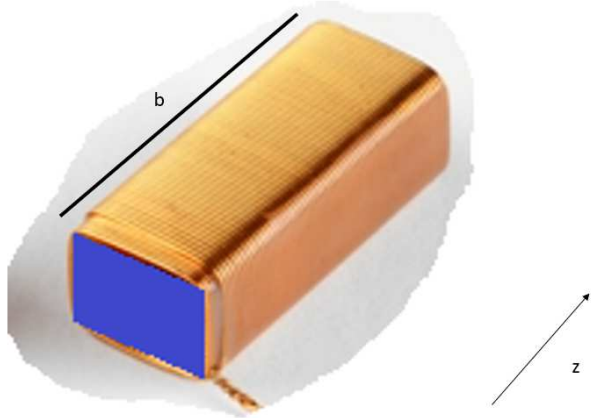}
\caption {A relativistic engine.}
 \label{releng1}
\end{figure}

\begin{figure}
\vspace{7cm} \includegraphics{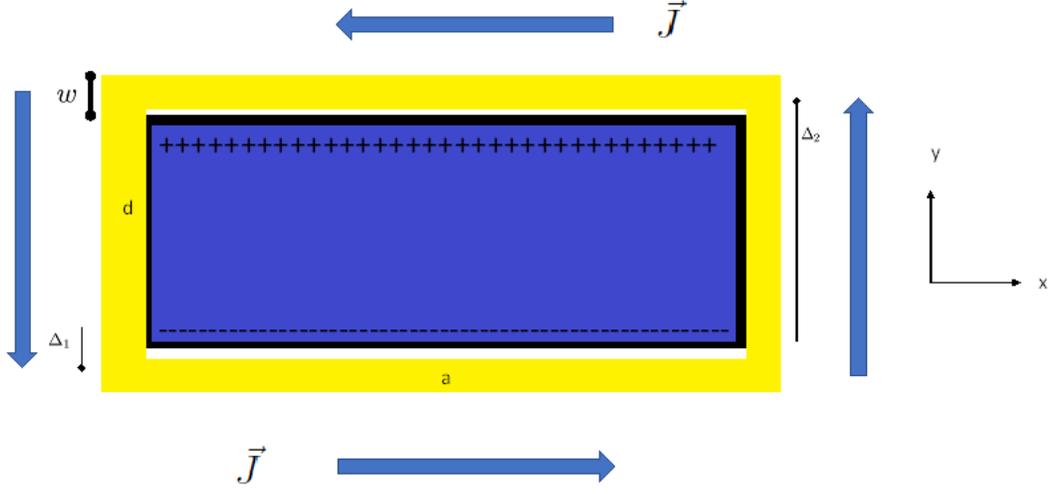}
\caption {A cross section of the relativistic engine.}
 \label{releng2}
\end{figure}

Consider an electret of thickness $d$ and area of $a \times b$, the electret is put inside a tight superconductive winded coil (see figures
\ref{releng1} and \ref{releng2}).
As the current in the $x$ direction is close to the negative charge it will make a considerable contribution to
the momentum equation \ern{P4} while the return current which is far away from the negative charge will make a relatively small contribution.
The situation is reversed with regard to the positive charge, here the return current will make the main contribution. Hence the total momentum
contribution is doubled with respect to the current flowing in the $x$ direction and interacting with negative charge alone. The current flowing
in the $y$ direction will not contribute to the momentum as the contribution from the current flowing upwards is exactly balanced by the contribution of the current flowing downwards. Let us look at the negative charge layer in figure \ref{releng2}, we shall model this as a surface charged layer in the plane $y=0$ with a surface charge density $\sigma$ such that:
\beq
\rho_1 (x_1, y_1,z_1) = - \sigma \delta (y_1) \left\{ \begin{array}{cc}
                                                       1 & -\frac{a}{2}<x_1<\frac{a}{2}, \quad  -\frac{b}{2}<z_1<\frac{b}{2} \\
                                                       0 & {\rm else}
                                                     \end{array}
    \right.
\label{schd}
\enq
in the above $\delta (y_1)$ is a Kronecker delta.
\beq
\vec J_2 (x_2, y_2,z_2) = \hat x J_0 w \delta (y_2+\Delta) \left\{ \begin{array}{cc}
                                                       1 & -\frac{a}{2}<x_2<\frac{a}{2}, \quad  -\frac{b}{2}<z_2<\frac{b}{2} \\
                                                       0 & {\rm else}
                                                     \end{array}
    \right.
\label{cude}
\enq
where $J_0$ is the current density, $w$ is the width of the winding, $\hat x$ is a unit vector in the $x$ direction and $\Delta$ is the distance of the current plane from the negative charge plane. Obviously their is a difference between the  $\Delta$ of the current: $\Delta_1$  and the return current: $\Delta_2$. The integrated current flowing through the system is:
\beq
I =  J_0 \ w \ b .
   \label{cur1}
\enq
However, the current flowing through each single wire depends on the winding number $N_w$:
\beq
I_c =  \frac{I}{N_w}
   \label{cur2}
\enq
Plugging \ern{schd} and \ern{cude} into \ern{P4}, we arrive at the following momentum equation:
\beq
\vec P = \frac{\mu_0}{4 \pi} \sigma J_0 w  \hat x \int_{-\frac{a}{2}}^{+\frac{a}{2}} dx_1 \int_{-\frac{b}{2}}^{+\frac{b}{2}} dz_1
  \int_{-\frac{a}{2}}^{+\frac{a}{2}} dx_2 \int_{-\frac{b}{2}}^{+\frac{b}{2}} dz_2 \frac{1}{R}.
\label{momentum1}
\enq
Such that:
\beq
R = \sqrt{(x_1-x_2)^2 + (z_1-z_2)^2 + \Delta^2}.
\label{Rexp}
\enq
Defining the dimensionless variables:
\beq
x'_1 = \frac{x_1}{a}, \quad z'_1 = \frac{z_1}{b}, \quad x'_2 = \frac{x_2}{a}, \quad z'_2 = \frac{z_2}{b}, \quad R' = \frac{R}{\Delta}, \quad a' = \frac{a}{\Delta}, \quad b' = \frac{b}{\Delta}.
\label{dimle}
\enq
We may write:
\beq
\vec P = \frac{\mu_0}{4 \pi} \sigma J_0  w a b^2 \tilde \Lambda(a',b') \hat x .
\label{momentum2}
\enq
The function $ \tilde \Lambda(a',b')$ is dimensionless and depends on the dimensionless quantities $a',b'$. It can be evaluated using
the quadruple integral:
\beq
\tilde \Lambda(a',b') \equiv \int_{-\frac{1}{2}}^{+\frac{1}{2}} dx'_1 \int_{-\frac{1}{2}}^{+\frac{1}{2}} dz'_1
  \int_{-\frac{1}{2}}^{+\frac{1}{2}} dx'_2 \int_{-\frac{1}{2}}^{+\frac{1}{2}} dz'_2 \frac{a'}{R'} .
\label{Lam}
\enq
In which:
\beq
R'= \sqrt{a'^2 (x'_1-x'_2)^2 + b'^2 (z'_1-z'_2)^2 + 1}.
\label{Rp}
\enq
We shall attempt to perform at least part of this integration analytically. For this, we shall define the auxiliary function:
\ber
\tilde \Gamma (a',b',x'_2,z'_2) &\equiv& \int_{-\frac{1}{2}}^{+\frac{1}{2}} dx'_1 \int_{-\frac{1}{2}}^{+\frac{1}{2}} dz'_1 \frac{a'}{R'}
\nonumber \\
\Rightarrow \tilde \Lambda(a',b') &=& \int_{-\frac{1}{2}}^{+\frac{1}{2}} dx'_2 \int_{-\frac{1}{2}}^{+\frac{1}{2}} dz'_2 \tilde \Gamma (a',b',x'_2,z'_2) .
\label{Gam}
\enr
We shall now  make a change of variables:
\beq
x= x'_1-x'_2, \qquad z=z'_1-z'_2,
\label{xz}
\enq
such that:
\ber
\tilde \Gamma (a',b',x'_2,z'_2) &=& \int_{-\frac{1}{2}-x'_2}^{+\frac{1}{2}-x'_2} dx
\int_{-\frac{1}{2}-z'_2}^{+\frac{1}{2}-z'_2} dz \frac{a'}{R'}
\nonumber \\
R' &=&  \sqrt{a'^2 x^2 + b'^2 z^2 + 1} .
\label{Gam2}
\enr
Finally, we define:
\ber
\Psi (a',b',x,z'_2) &\equiv& \int_{-\frac{1}{2}-z'_2}^{+\frac{1}{2}-z'_2} dz \frac{a'}{R'}
\nonumber \\
\Rightarrow \tilde \Gamma (a',b',x'_2,z'_2) &=& \int_{-\frac{1}{2}-x'_2}^{+\frac{1}{2}-x'_2} dx \Psi (a',b',x,z'_2) .
\label{Psi}
\enr
The function $\Psi$ can be integrated analytically by introducing the new variable:
\beq
z'' = \frac{b}{a} \frac{z}{\sqrt{x^2+a'^{-2}}}
\label{al}
\enq
in terms of which:
\ber
\Psi (a',b',x,z'_2) &=&  \frac{a}{b}  \int_{z''_1}^{z''_2} dz'' \frac{1}{\sqrt{1+z''^2}}
=\frac{a}{b} \left[{\rm arcsinh} (z''_2) - {\rm arcsinh} (z''_1) \right]
\nonumber \\
z''_1 &=& \frac{b}{a} \left(\frac{-\frac{1}{2}-z'_2}{\sqrt{x^2+a'^{-2}}}\right), \qquad
z''_2 = \frac{b}{a} \left(\frac{+\frac{1}{2}-z'_2}{\sqrt{x^2+a'^{-2}}}\right).
\label{Psi2}
\enr
Unfortunately the calculation of $\tilde \Gamma$ and $\tilde \Lambda$ can only be done numerically.

\begin{figure}
\vspace{7cm} \includegraphics{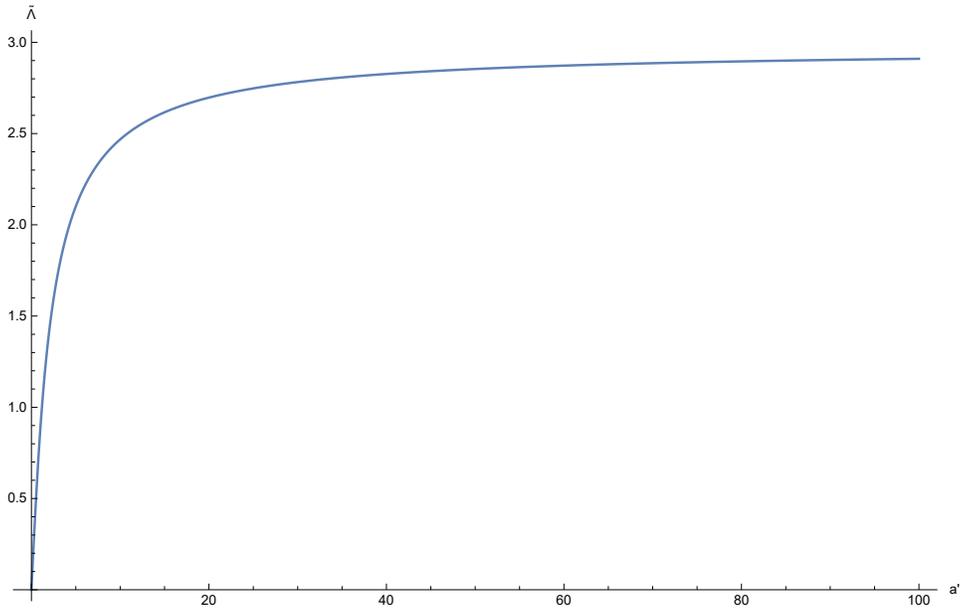}
\caption {The function $\tilde \Lambda$ for the case $b'=a'$.}
 \label{lam}
\end{figure}

\begin{figure}
\vspace{7cm} \includegraphics{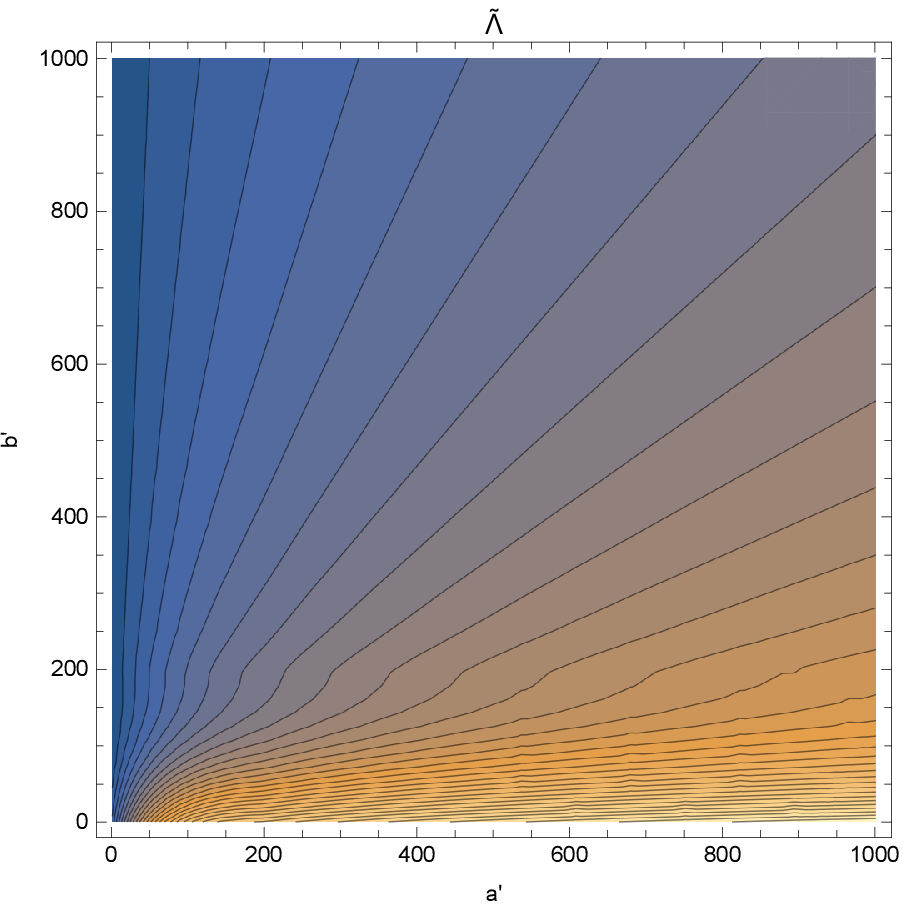}
\includegraphics{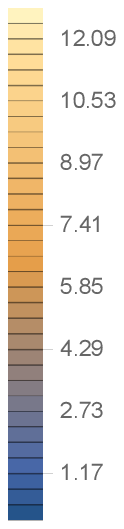}
\caption {The function $\tilde \Lambda$ for the case $b'\neq a'$, the color code is shown to the right of the contour plot.}
 \label{2Dgr}
\end{figure}

For the case $b'=a'$ the function $\tilde \Lambda$ is a single variable function depicted in figure \ref{lam}.
It can be seen that for $a'=0$ the function is null, while for $a'=\infty$ it approaches $2.973 \sim 3$. Hence if the ratio of
 size of the engine  to the winding width is about $20$ and $d \sim a$ one does not loss a significant amount of force \& momentum due
 to the return current. However, looking at the case $b'\neq a'$ depicted in figure \ref{2Dgr}, we a see an obvious advantage for a slender  relativistic motor in the direction of motion.
We also point out that the return current interacts symmetrically with the positive charge in a beneficial way, as both the sign of the charge
and the direction of the current are reversed (see figure \ref{releng2}) this doubles the momentum gain such that according to \ern{momentum2} we have:
\beq
\vec P_T = \frac{\mu_0}{2 \pi} \sigma J_0  w a b^2  \left(\tilde \Lambda(\frac{a}{\Delta_1},\frac{b}{\Delta_1}) -
\tilde \Lambda(\frac{a}{\Delta_2},\frac{b}{\Delta_2})\right) \hat x .
\label{momentumT}
\enq
The quantities $\Delta_1$ and $\Delta_2$ are defined through figure \ref{releng2} and we thus expect:
\beq
\Delta_1 = \frac{w}{2}, \qquad \Delta_2 = \frac{w}{2}+d.
\label{del12}
\enq
The force produced by the relativistic engine depends on the rise time of the current which can be increased gently or abruptly:
\beq
\vec F_T = \frac{d \vec P_T}{dt}= \frac{\mu_0}{2 \pi} \sigma \frac{d J_0}{dt}  w a b^2  \left(\tilde \Lambda(\frac{a}{\Delta_1},\frac{b}{\Delta_1}) - \tilde \Lambda(\frac{a}{\Delta_2},\frac{b}{\Delta_2})\right) \hat x .
\label{FT}
\enq
We choose three different configurations of a relativistic motor, one is a size of standard car and its parameters are described in
the first column of table \ref{partabb} the other describes a considerably larger motor which may be suitable for space travel is given in
the second column of table \ref{partabb}, the third column describes a cube of immense dimensions.
\begin{table}
\begin{center}
\begin{tabular}{|c|c|c|c|c|}
  \hline
           &car &  rocket size engine & giant cube & units \\
  \hline
  $a$ & 6 & 200 & 1000 & m \\
  $b$ & 2 & 10 & 1000 & m \\
  $d$ & 1 & 10 & 1000 & m \\
  $w$ & 0.2 & 0.4 & 0.4 & m \\
  \hline
  $P_T$ & 0.3 & 868 & $3.1 \ 10^7$  & kg m/s \\
  \hline
\end{tabular}
\end{center}
\caption{Maximal momentum gained by a relativistic motor for three cases of parameters. We assume an extreme case of charge density $\sigma = 3.7 \ 10^ {-3}$ Coulomb/$\rm m^2$ (see section \ref{chden}), and current density $J_0 = 5 \ 10^{7}$  Ampere/$\rm m^2$ (see section \ref{cuden}).}
\label{partabb}
\end{table}

We see that despite the extreme parameters the maximum momenta gained are quite modest,this is a direct result of the phenomena of dielectric breakdown.

\section{The nano relativistic motor}

\begin{figure}
\vspace{7cm} \includegraphics{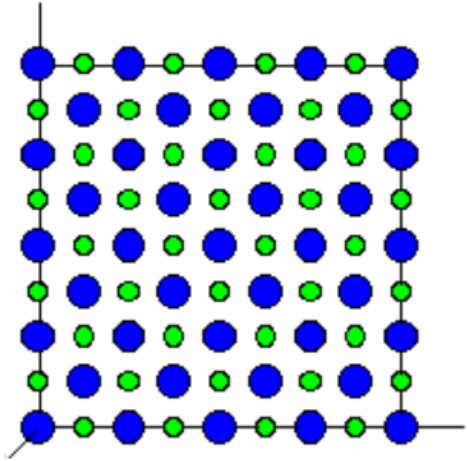}
\caption {The $100$ plane of a table salt $Na^+ Cl^-$ lattice, blue circles stand for Sodium positive ions and yellow circles stand for Chlorine negative ions.}
 \label{Nacl}
\end{figure}

In the previous section we have shown that intrinsic parameter limitations, especially dielectric breakdown lead to somewhat modest amount of momenta that can be gained by relativistic motor. However, it seams that in the microscopic domain this limitation is rather relaxed. For example if consider ionic crystals such as the prevalent table salt: $Na^+ Cl^-$. This salt solidifies to a face centered cubic lattice in which the lattice constant is $l=564.02$ pm. Taking for example the $100$ plane of this lattice (see figure \ref{Nacl}) we can see that the charge density is periodical in which in each half unit cell we have a surface charge density of $\pm  2~ {\rm Coulomb/m^2}$. This is of course thousand time larger than available macroscopic charge densities (see section \ref{chden}), however, on the macroscopic average this leads to a null charge density.
To see how we can circumvent this unfortunate reality we will investigate \ern{P4}, calculating the spatial Fourier transform of the scalar potential and the current density such that:
\ber
\Phi_1 (\vec k) &=& \int_{-\infty}^{+\infty} \int_{-\infty}^{+\infty} \int_{-\infty}^{+\infty} \Phi_1 (\vec x) e^{-i \vec k \cdot
\vec x} d^3 x
\nonumber \\
\vec J_2 (\vec k) &=& \int_{-\infty}^{+\infty} \int_{-\infty}^{+\infty} \int_{-\infty}^{+\infty} \vec J_2  (\vec x) e^{-i \vec k \cdot
\vec x} d^3 x.
\label{fourtrans}
\enr
Using the theorem of Parseval \cite{peebles}, we may now write \ern{P4} in the form:
\beq
\vec P (t)= - \frac{1}{c^2 (2 \pi)^3} \int_{-\infty}^{+\infty} \int_{-\infty}^{+\infty} \int_{-\infty}^{+\infty}   d^3 k ~ \Phi_1^* (\vec k)
 \vec J_2 (\vec k)
\label{P5}
\enq
In this form it is obvious, that a microscopic distribution of periodic charge density will be beneficial if it is a accompanied by a current density distribution of the same period. How can we obtain such a microscopic charge density distribution? To this end we remind the reader that
microscopic currents are associated with the electronic motion and electronic spin. The magnetization $\vec{M}$ is related to the magnetization current $\vec{J}_M$ by the formulae \cite{MTAY3,AY1}:
\beq
\vec{J}_M \equiv \curl{\vec{M}}.
\label{JM}
\enq
We may replace in \ern{P3b}, \ern{P4} and \ern{P5} $\vec{J}$ by $\vec{J}_M$ to obtain a similar effect. Furthermore, the magnetization $\vec{M}$ is related to a the microscopic dipole moments $\vec m_i$ through:
\beq
\vec{M} \equiv \frac{1}{V} \sum_i \vec m_i.
\label{M}
\enq
in which we sum over all dipoles and divide by the sample volume $V$. In known magnetic materials such as iron the magnetic dipole moments are related to the spin configuration of the atom. In  $\alpha$-iron the spins of the two unpaired electrons in each atom generally align with the spins of its neighbors. This happens because the orbitals of those two electrons ($d_{z^2}$ and $d_{x^2 - y^2}$) do not point toward neighboring atoms in the lattice, and therefore are not involved in metallic bonding. Hence the ideal configuration for a relativistic motor will involve and ionic lattice in which one species of atom (say the positively charged) will involve free spins that can be manipulated by an external magnetic field thus creating a relativistic engine effect which can be easily manipulated in three axis. The details of such an engine are beyond the scope of the current paper.

\section{Conclusion}

In this paper, we have shown that, in general, Newton's third law is not compatible with the principles of special relativity and the total force on a two charged body system is not zero. Still, momentum is conserved if one takes the field momentum into account, and the same is true for energy.

The main results of this paper are given by \ern{P} and \ern{P2}, which describe the total relativistic force for charged bodies. Not all configurations of charged relativistic motors were analyzed here due to space limitations. We have concentrated our efforts on a system in which one part is charged and the second part carries current. It is noticed that the charge relativistic motor is many orders of magnitude powerful than an uncharged relativistic motor. Still, it is shown that the limitations of dielectric breakdown and current density lead to a rather modest momentum generated by the engine on the macroscopic level.

Those limitations can be somewhat relaxed at the atomic level, as the surface charge density for typical ionic crystals is three orders of magnitude larger than what can be achieved at the macroscopic level. Thus future work will concentrate on the development of nano relativistic engines. Other open subjects are exploring additional possible charged relativistic configurations as suggested by \ern{P2} and studying in detail momentum and energy transformation from the electromagnetic to the mechanical elements of the motor.

\vfill \eject
\noindent {\Huge  \bf Appendices}

\appendix

\section {The consistency of the gauge condition}

Set $\vec x = \vec x_2$, $\vec x' = \vec x_1$ and $R =|\vec x' - \vec x|$, thus \ern{A} takes the form:
\beq
\vec A (\vec x,t) = \frac{\mu_0}{4 \pi} \int d^3 x' \frac{\vec J  (\vec x',t_{ret})}{R}.
\label{A22}
\enq
in which we suppressed the subscript of $\vec J $. Taking a divergence of both sides of the above equation we obtain:
\ber
\vec \nabla \cdot \vec A  &=& \frac{\mu_0}{4 \pi} \int d^3 x' \vec \nabla \cdot \left(\frac{\vec J  (\vec x',t_{ret})}{R}\right)
\nonumber \\
&=& \frac{\mu_0}{4 \pi} \int d^3 x'  \left(\frac{\vec \nabla \cdot \vec J  (\vec x',t_{ret})}{R} +
 \vec J (\vec x',t_{ret}) \cdot \vec \nabla  \frac{1}{R} \right).
\label{A22d}
\enr
However:
\beq
\vec \nabla  \frac{1}{R}  = - \vec \nabla'  \frac{1}{R}
\label{vecnap}
\enq
Hence:
\beq
\vec \nabla \cdot \vec A  = \frac{\mu_0}{4 \pi} \int d^3 x'  \left(\frac{\vec \nabla \cdot \vec J  (\vec x',t_{ret})}{R} -
 \vec J (\vec x',t_{ret}) \cdot \vec \nabla'  \frac{1}{R} \right).
\label{A22d2}
\enq
This leads to:
\beq
\vec \nabla \cdot \vec A  = \frac{\mu_0}{4 \pi}  \left(\int d^3 x' \frac{\vec \nabla \cdot \vec J  (\vec x',t_{ret})+\vec \nabla' \cdot \vec J  (\vec x',t_{ret})}{R} - \oint d \vec S \cdot  \frac{\vec J (\vec x',t_{ret})}{R} \right) .
\label{A22d3}
\enq
in which we have used Gauss theorem. The surface integral is taken over a closed surface encapsulating the volume of integration. However, the volume can be infinite and thus the surface is taken at infinity at which we assume there are no current densities. Hence:
\beq
\vec \nabla \cdot \vec A  = \frac{\mu_0}{4 \pi} \int d^3 x' \frac{\vec \nabla \cdot \vec J  (\vec x',t_{ret})+\vec \nabla' \cdot \vec J  (\vec x',t_{ret})}{R} .
\label{A22d4}
\enq
now $\vec J (\vec x',t_{ret})$ depends on $\vec x'$ both directly and through $t_{ret}$, thus:
\beq
\vec \nabla' \cdot \vec J  (\vec x',t_{ret}) = \vec \nabla' |_{t_{ret}} \cdot \vec J  (\vec x',t_{ret}) + \partial _{t_{ret}} \vec J  (\vec x',t_{ret}) \cdot \vec \nabla' t_{ret}.
\label{A22d5}
\enq
$\vec \nabla' |_{t_{ret}} \cdot \vec J  (\vec x',t_{ret})$ means taking a divergence with respect to $\vec x'$ but leaving $t_{ret}$ constant.
However:
\beq
 \vec \nabla' t_{ret} = \vec \nabla' R ~\partial_R t_{ret} =  - \vec \nabla R ~\partial_R t_{ret} = -  \vec \nabla t_{ret}
\label{A22d6}
\enq
Hence:
\beq
\vec \nabla' \cdot \vec J  (\vec x',t_{ret}) = \vec \nabla' |_{t_{ret}} \cdot \vec J  (\vec x',t_{ret}) - \partial _{t_{ret}} \vec J  (\vec x',t_{ret}) \cdot \vec \nabla t_{ret}.
\label{A22d7}
\enq
We notice that:
\beq
\vec \nabla \cdot \vec J  (\vec x',t_{ret}) = \partial _{t_{ret}} \vec J  (\vec x',t_{ret}) \cdot \vec \nabla t_{ret},
\label{A22d8}
\enq
it now follows:
\beq
\vec \nabla' \cdot \vec J  (\vec x',t_{ret}) = \vec \nabla' |_{t_{ret}} \cdot \vec J  (\vec x',t_{ret}) -\vec \nabla \cdot \vec J  (\vec x',t_{ret}).
\label{A22d9}
\enq
Inserting \ern{A22d9} into \ern{A22d4} leads to:
\beq
\vec \nabla \cdot \vec A  = \frac{\mu_0}{4 \pi} \int d^3 x' \frac{\vec \nabla' |_{t_{ret}} \cdot \vec J  (\vec x',t_{ret})}{R} .
\label{A22d10}
\enq
Let us now look at the scalar potential given in \ern{Phi}, this can be written using the current notation as:
\beq
\Phi =  \frac{1}{4 \pi \epsilon_0} \int d^3 x' \frac{\rho (\vec x',t_{ret})}{R}.
\label{PhiA2}
\enq
Taking a partial temporal derivative of $\Phi$ and multiplying by $\frac{1}{c^2} = \epsilon_0 \mu_0$ we arrive at:
\beq
\frac{1}{c^2} \partial_t \Phi =  \frac{\mu_0}{4 \pi} \int d^3 x' \frac{\partial_t \rho (\vec x',t_{ret})}{R}.
\label{PhiA3}
\enq
Combining \ern{PhiA3} with \ern{A22d10} it follows that:
\beq
\frac{1}{c^2} \partial_t \Phi + \vec \nabla \cdot \vec A  =  \frac{\mu_0}{4 \pi} \int d^3 x'
\left.\frac{\partial_t \rho (\vec x',t) + \vec \nabla'  \cdot \vec J  (\vec x',t)}{R}\right|_{t = t_{ret}}.
\label{PhiA4}
\enq
However, due to the charge conservation \ern{Chacon} we obtain:
\beq
\frac{1}{c^2} \partial_t \Phi + \vec \nabla \cdot \vec A  =  0.
\label{PhiA5}
\enq


\begin {thebibliography} {99}

\bibitem {MTAY1}
Miron Tuval \& Asher Yahalom "Newton's Third Law in the Framework of Special Relativity" Eur. Phys. J. Plus (11 Nov 2014) 129: 240
 DOI: 10.1140/epjp/i2014-14240-x. (arXiv:1302.2537 [physics.gen-ph]).
\bibitem {MTAY3}
Miron Tuval and Asher Yahalom "A Permanent Magnet Relativistic Engine" Proceedings of the Ninth International Conference on  Materials
Technologies and Modeling (MMT-2016) Ariel University, Ariel, Israel, July 25-29, 2016.
\bibitem {AY1}
Asher Yahalom "Retardation in Special Relativity and the Design of a Relativistic Motor". Acta Physica Polonica A, Vol. 131 (2017) No. 5, 1285-1288. DOI: 10.12693/APhysPolA.131.1285
\bibitem {MTAY4}
Miron Tuval and Asher Yahalom "Momentum Conservation in a Relativistic Engine" Eur. Phys. J. Plus (2016) 131: 374. \\ DOI: 10.1140/epjp/i2016-16374-1
\bibitem {AY2}
Asher Yahalom "Preliminary Energy Considerations in a Relativistic Engine" Proceedings of the Israeli-Russian Bi-National Workshop "The optimization of composition, structure and properties of metals, oxides, composites, nano - and amorphous materials", page 203-213, 28 - 31 August 2017, Ariel, Israel.
\bibitem {RY}
S. Rajput and A. Yahalom, "Preliminary Magnetic Energy Considerations in a Relativistic Engine: Mutual Inductance vs. Kinetic Terms" 2018 IEEE International Conference on the Science of Electrical Engineering in Israel (ICSEE), Eilat, Israel, 2018, pp. 1-5. doi: 10.1109/ICSEE.2018.8646265
\bibitem {RY2}
S. Rajput and A. Yahalom, "Electromagnetic Radiation of a Relativistic Engine: Preliminary Analysis" accepted for publication in the proceedings of the International Congress on Advanced Materials Sciences and Engineering 22-24 July, 2019, Osaka, Japan.
\bibitem {Einstein}
A. Einstein, "On the Electrodynamics of Moving Bodies", Annalen der Physik 17 (10): 891�921, (1905).
\bibitem {Maxwell}
 J.C. Maxwell, "A dynamical theory of the electromagnetic field"
  Philosophical Transactions of the Royal Society of London 155: 459�512 (1865).
\bibitem {Jackson}
J. D. Jackson\index{Jackson J.D.}, Classical Electrodynamics\index{electrodynamics!classical}, Third Edition. Wiley: New York, (1999).
\bibitem {Feynman}
R. P. Feynman, R. B. Leighton \& M. L. Sands, Feynman Lectures on Physics, Basic Books; revised 50th anniversary edition (2011).
\bibitem {Heaviside}
O. Heaviside, "On the Electromagnetic Effects due to the Motion of Electrification through a Dielectric" Philosophical Magazine, (1889).
\bibitem {Newton}
I. Newton, Philosophiae Naturalis Principia Mathematica (1687).
\bibitem {Goldstein}
H. Goldstein , C. P. Poole Jr. \& J. L. Safko, Classical Mechanics, Pearson; 3 edition (2001).
\bibitem {Mansuripur}
M. Mansuripur, "Trouble with the Lorentz Law of Force: Incompatibility with Special Relativity and Momentum Conservation" PRL 108, 193901 (2012).
\bibitem {Griffiths}
D. J. Griffiths \&  M. A. Heald, "Time dependent generalizations of the Biot-Savart and Coulomb laws"
American Journal of Physics, 59, 111-117 (1991), DOI:http://dx.doi.org/10.1119/1.16589
\bibitem {Jefimenko}
Jefimenko, O. D., Electricity and Magnetism, Appleton-Century Crofts, New York (1966); 2nd edition, Electret Scientific, Star City, WV (1989).
\bibitem {factbook}
 Dielectric Strength of Air - the Physics Fact book\\ https://hypertextbook.com/facts/2000/AliceHong.shtml
\bibitem {vacuumds}
Stefan Giere, Michael Kurrat \& Ulf Sch\"{u}mann, HV Dielectric Strength of Shielding Electrodes in Vacuum Circuit-Breakers,
$20^{th}$ International Symposium on Discharges and Electrical Insulation in Vacuum - Tours, France - June 30 - July 5, 2002
\bibitem {diamondds}
Markus Gabrysch "Electronic properties of diamond".\\ el.angstrom.uu.se. Retrieved 2013-08-10.
\bibitem {Hsi}
L. Hsi-wen, T. Yu-Chong, Parylene-based electret power generators, J. Micromech. Microeng. 18 (10)
(2008) 104006.
\bibitem {Jung}
Jung, S.-G. et al. Enhanced critical current density in the pressure-induced
magnetic state of the high-temperature superconductor FeSe. Sci. Rep. 5, 16385; doi: 10.1038/srep16385 (2015).
\bibitem {peebles}
Peebles, P.~Z. Probability, Random Variables and Random Signal Principles, McGraw Hill, New York, NY, USA (2001).
\end{thebibliography}
\end{document}